\documentstyle[twocolumn,psfig,rotating]{mn}
\def\pmb#1{\setbox0=\hbox{#1}%
\kern-.025em\copy0\kern-\wd0
\kern.05em\copy0\kern-\wd0
\kern-.025em\raise.0433em\box0}

\newcommand{\kmsmpc}{\ {\rm km\,s^{-1}\,Mpc^{-1}}}
\newcommand{\hmpc}{\ {\rm {\it h}^{-1}Mpc}}
\newcommand{\mdh}{\ {\rm M_\odot/{\it h}}}
\def\om#1{\Omega_{\rm #1}}
\newcommand\lsim{~\rlap{$<$}{\lower 1.0ex\hbox{$\sim$}}}
\newcommand\gsim{~\rlap{$>$}{\lower 1.0ex\hbox{$\sim$}}}

\newcommand{\mhr}{m_{_{\rm HR}}}  
\newcommand{\xm}{\xi_{_{\rm mm}}} 
\newcommand{\xh}{\xi_{_{\rm hh}}} 

\newcommand{\etal}{{\it et al.\ }}

\def\dd{{\rm d}}
\def\vx{{\bf x}}
\def\vg{{\bf g}}
\def\vnabla{{\bf \nabla}}
\def\dm{\delta_{\rm m}}
\def\dh{\delta_{\rm h}}
\begin{document}
\title[] {Numerical action reconstruction of the dynamical history of 
dark matter haloes in $N$-body simulations}
\author[Phelps, Desjacques, Nusser, and Shaya]{Steven D. Phelps$^{1,2}$, 
Vincent Desjacques$^{1\star}$, Adi Nusser$^1$, and Edward J. Shaya$^3$ \\
$^1$ Physics Department, The Technion, Haifa 32000, Israel \\
$^2$ Bah\'{a}'\'{i} World Centre, Haifa 31001, Israel \\
$^3$ Astronomy Department, University of Maryland, College Park, MD 20743 \\
Email~: phelps@bwc.org\\}
\maketitle

\begin{abstract}

We test the ability of the numerical action method (NAM) to recover
the individual orbit histories of mass tracers in an expanding
universe in a region of radius $26\hmpc$, given the masses and
redshift-space coordinates at the present epoch.  The mass tracers are
represented by dark matter haloes identified in a high resolution
$N$-body simulation of the standard $\Lambda$CDM cosmology.  Since
previous tests of NAM at this scale have traced the underlying
distribution of dark matter particles rather than extended haloes, our
study offers an assessment of the accuracy of NAM in a scenario which
more closely approximates the complex dynamics of actual galaxy
haloes.  We show that NAM can recover present-day halo distances with
typical errors of less than 3 per cent, compared to 5 per cent errors
assuming Hubble flow distances.  The total halo mass and the linear
bias were both found to be constained at the 50 per cent level.  The
accuracy of individual orbit reconstructions was limited by the
inability of NAM, in some instances, to correctly model the positions
of haloes at early times solely on the basis of the redshifts, angular
positions, and masses of the haloes at the present epoch.  
Improvements in the quality of NAM reconstructions may be possible
using the present-day three-dimensional halo velocities and distances
to further constrain the dynamics.  This velocity data is expected to
become available for nearby galaxies in the coming generations of
observations by SIM and GAIA.
 
\end{abstract}

\begin{keywords}
cosmology: theory -- galaxies : distances and redshifts
\end{keywords}

\section{Introduction}

In the standard cosmological paradigm, the formation of large-scale
structure is driven by the gravitational amplification of small
initial density fluctuations (e.g. Peebles 1980).  In addition to
gravity, hydrodynamical processes can influence the formation and
evolution of galaxies, groups and clusters of galaxies.  But since
hydrodynamical effects play a minor role on scales larger than the
size of galaxy clusters, gravitational instability theory alone can
directly relate the present day large-scale structure to the initial
density field and provide the framework within which the observations
can be analyzed and interpreted.  Gravitational instability is a
nonlinear process, making numerical methods an essential tool for
understanding the observed large-scale structure.~\footnote{
Present address~: Racah Institute of Physics, The Hebrew University, 
Jerusalem Israel}

There are two complementary numerical approaches to studying
cosmological structure.  The first relies on $N$-body techniques
designed to solve an initial value problem in which the evolution of a
self-gravitating system of massive particles is determined by forward
numerical integration of the Newtonian differential equations.
Because the exact initial conditions are unknown, comparisons between
these simulations and observations are mainly concerned with general
statistical properties.

The second approach works in the opposite direction, deriving from the
observed present-day distribution and peculiar motions of galaxies,
and independently of the nature of the dark matter, certain features
of the dynamics at earlier times.  The numerical action method (NAM)
belongs to this second category of approaches.  It arises from the
observation that the present-day distribution of galaxies, combined
with the reasonable assumption that their peculiar velocities vanish
at early times, presents a boundary value problem that naturally lends
itself to an application of Hamilton's principle in which stationary
variations of the action are found subject to the boundary conditions.
The result is a prediction of the full orbit histories of individual
galaxies, either with real space boundary conditions (Peebles 1989,
1990, 1994, 1995) or, after a coordinate transformation, in redshift
space (Peebles \etal 2001, Phelps 2002).

The potential of NAM as a probe of galaxy dynamics and of cosmological
parameters has been explored in a number of studies following the
introduction of the method in Peebles 1989.  Possible applications
include the full nonlinear analysis of orbit histories of nearby
galaxies (Peebles 1990, 1994, 1995; Sharpe \etal 2001), recovering the
initial power spectrum of density fluctuations (Peebles 1996),
predicting the values of cosmological parameters (Shaya \etal 1995),
and estimating the proper motions of nearby galaxies (Peebles \etal
2000).  Concerning the latter application, ground and space-based
observations will soon make possible the measurement of the full
three-dimensional velocities of many nearby galaxies and promise both
a rigourous test of NAM predictions and, given the additional
dynamical constraints on galaxy motions, the possiblity of using NAM
as a probe of individual masses of nearby galaxies.

Since a central result of NAM, the past orbit histories of galaxies,
cannot be confirmed by direct observations, $N$-body simulations
provide an important test of NAM and its key assumption that
galaxies can be approximated as discrete, non-merging objects
throughout their history.  It is desirable then to test NAM in a
scenario which approximates the complexity of the observational
situation but where all of the relevant physical quantities are known.
Previous tests of NAM using $N$-body simulations have either been
confined to a few dark matter haloes at the scale of the Local Group
(Branchini \& Carlberg 1994, Dunn \& Laflamme 1995), traced the paths
of individual dark matter particles rather than extended haloes
(Nusser \& Branchini 2000), or used simulations which demonstrate in
principal the ability of NAM to recover particle orbits to a high
degree of accuracy but which do not reproduce the full complexity of
extended mass distributions (Phelps 2002).

In this paper we extend the tests of NAM to simulations at a scale
approaching that of the local supercluster with a catalogue containing
several hundred extended objects modelled as particles.  We begin with
an overview of the relevant properties of the $N$-body simulation and
the halo catalogue we derived from it, and follow with details of the
version of NAM used here, which includes a novel approach to the
assignment of halo masses.  We will then test the sensitivity of NAM
both as a probe of the total mass as well as of the linear bias, and
examine in some detail a representative solution, focusing on the
comparison between the NAM predictions and the actual halo orbits.

\section{The simulation}

\begin{figure}
\resizebox{0.48\textwidth}{!}{\includegraphics{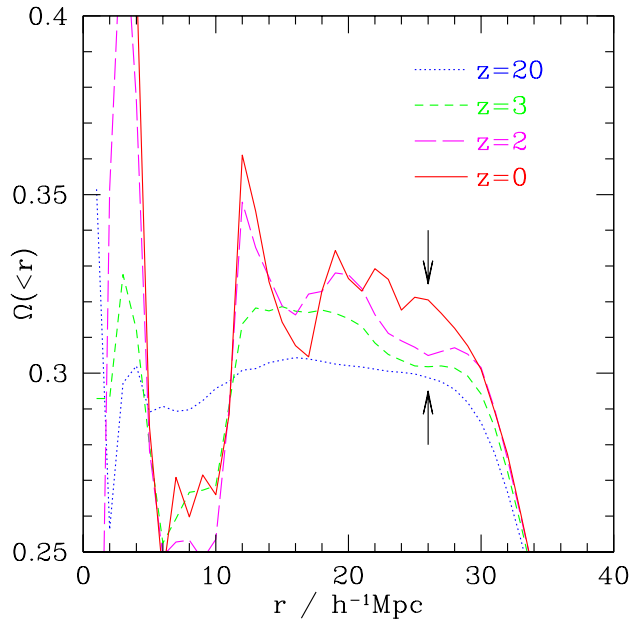}}
\caption{The average matter density $\Omega(<r)$ in a sphere of radius
$r$ from the center of mass of the simulation, as a function of
redshift. The arrows mark the nominal initial boundary ($r=26\hmpc$)
of the high-resolution region. The average density within $26\hmpc$ is
about 10 per cent larger than the cosmic mean $\om{m}=0.3$ at $z=0$.}
\label{fig:omega}
\end{figure}

We will test NAM in a $N$-body simulation of a $\Lambda$CDM cosmology
with matter density $\om{m}=0.3$, vacuum energy density
$\om{\Lambda}=0.7$ (in units of the critical density $\rho_{\rm c}$),
and a Hubble constant $h=0.7$ (in units of 100$\kmsmpc$). The
simulation has an initial fluctuation power spectrum of spectral index
$n=1$ and a present-day normalization amplitude $\sigma_8=0.9$. It was
run with the GADGET $N$-body code (Springel, Yoshida \& White
2001). It is part of a suite of simulations which zoom in on a
spherical region of initial comoving radius 26$\hmpc$ selected from a
larger parent simulation (see Stoehr 2003 for details).  This
``zooming'' technique has the advantage of preserving the large-scale
power of the parent simulation while allowing the inner region to be
resolved at smaller scales.  In particular, the mass of the 168436
``high-resolution'' particles which sample the inner region is
$\mhr=6.82\times 10^{10}\mdh$, in contrast to the 602272
``low-resolution'' particles (each having a mass $\gg \mhr$) which
sample the outer region.

Fig.~\ref{fig:omega} shows the average matter density $\Omega(<r)$
averaged over a sphere of radius $r$ as a function of redshift.  At
the initial time step ($z=20$), the mean density in a sphere of radius
$r=26\hmpc$, $\Omega(<26)$, is very close to $\om{m}$, but increases
by about 10 per cent from initial to final time.  The net mass inflow
across the region boundary has the potential to disrupt the
reconstruction at the edge of the region and provides a further
opportunity to test the sensitivity of NAM to the total mass.

\begin{figure}
\resizebox{0.48\textwidth}{!}{\includegraphics{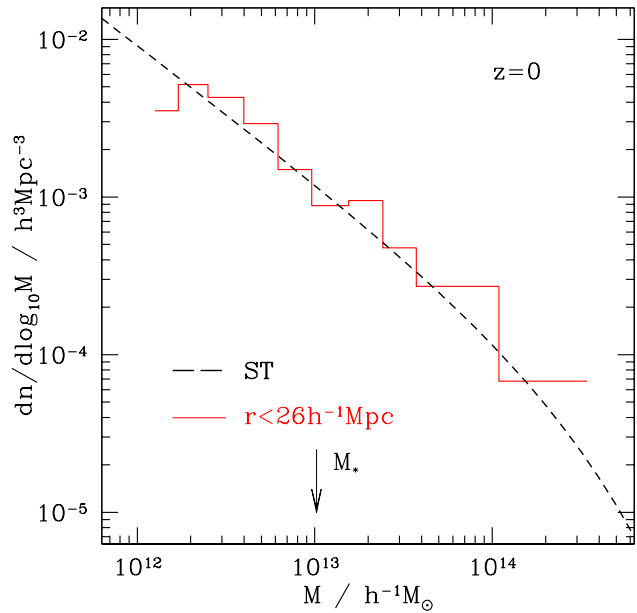}}
\caption{The friends-of-friends differential mass function of dark
matter haloes (in logarithmic bins $\Delta\log_{10}m=0.2$) computed
from the snapshot at $z=0$ (histogram) compared  to the theoretical
prediction of Sheth \& Tormen (dashed curve).  The arrow indicates
$M_\star\sim 10^{13}\mdh$, or the typical mass of haloes collapsing
at the present time. }
\label{fig:haloes}
\end{figure}

\subsection{The halo catalogue}

The principal objective of this paper is to test the ability of NAM to
reconstruct the orbits of simulated dark matter (DM) haloes which resemble
the galaxy + halo distributions of the ``real world''. To this end we
need to specify the positions, velocities and masses of a {\it fixed}
number of haloes as a function of time, since our implementation of NAM
conserves particle number.  However, haloes do not form
at any single redshift, and mergers potentially play an important role in the
dynamics. 

In order to create a catalogue of a fixed number of haloes from the
inner region of the simulation, we first identify all the DM haloes
above a given mass threshold $M_{\rm min}$ in the final simulation
time step $z=0$, using a friends-of-friends (FOF) group-finding
algorithm based on the method given in Davis \etal (1985). Only groups
containing at least 20 DM particles are classified as haloes,
resulting in a mass threshold $M_{\rm min}=1.36\times 10^{12}\mdh$. We
used the standard linking length of 20 per cent of the mean
interparticle distance (Jenkins \etal 2001).  Using haloes defined
with a linking length of 30 per cent we found a significant
improvement in the accuracy of NAM reconstructions, which is not
unexpected since this probes a more linear regime.  In this paper
however we confine our analysis to the standard linking length since
it most closely reproduces the expected mass function and thus gives
the best approximation to real galaxy haloes.

The center of mass (CM) positions and velocities of all those DM
particles which are associated with haloes at $z=0$ define the
resulting catalogue of 576 objects which is the input to the numerical
action code. The orbit histories of these objects are then defined as
the CM positions and velocities of the DM particles which belong to
the corresponding halo at $z=0$.  For our purposes the redshift
outputs at $z$=20, 4, 3, 2, 1 and 0 were sufficient for comparison of
halo orbits with the action reconstruction.

To model the tidal field from the outer region of the simulation, we
created an auxiliary particle catalogue from a random sample of 1000
low-resolution DM particles, multiplying the masses by the appropriate
factor of $277.9$ to bring the mass density in the outer region to
$\om{m} = .3$.  To reduce the computational burden, the positions of
these tidal field particles were fixed in the NAM reconstructions and
their masses were evolved according to linear perturbation theory, as
described in Shaya \etal (1995).  Parallel sets of NAM solutions were
created with and without this tidal field.

The final step is to identify one of the haloes near the center of the
simulation as the reference halo, and to build an input catalogue to
NAM using the angular positions and redshifts relative to this halo.
The analysis below was repeated for two haloes within 8 Mpc/h of the
CM of the high-resolution region, as a check on the sensitivity to the
position and velocity of the reference halo, and since the results
were consistent we will confine our analysis to a single reference halo.

In Fig.~\ref{fig:haloes} we show as a histogram the FOF differential
mass function of DM haloes in the high-resolution region at $z=0$. The
arrow marks $M_\star\sim 10^{13}\mdh$, the typical mass of haloes that have
just formed by redshift $z=0$. As can be seen, the halo masses span
more than two orders of magnitude, with the largest  haloes having a
mass $M\gsim 10^{14}\mdh$.  The dashed curve is the theoretical
prediction of Sheth \& Tormen (1999) for a $\Lambda$CDM cosmology
matching that of the simulation. The good agreement shows that the
high-resolution region of the simulation is a fair  reproduction of
the expected mass function.

\begin{figure}
\resizebox{0.48\textwidth}{!}{\includegraphics{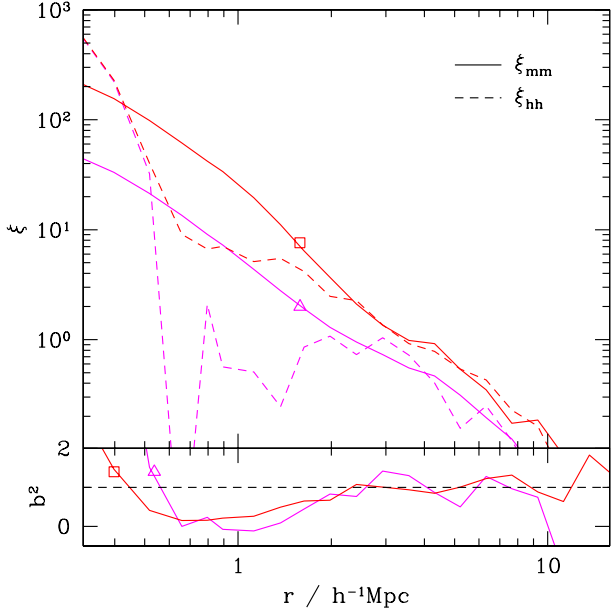}}
\caption{ The auto-correlation of haloes (dashed
curve) and dark matter (solid curve) obtained from the unsmoothed
halo and matter distribution at $z=0$ (square) and $z=2$ (triangle). 
The bias $b^2=\xh/\xm$ is plotted in the lower panel. The horizontal
dashed line shows $b=1$.}
\label{fig:bias}
\end{figure}

In the top panel of Fig.~\ref{fig:bias} we compare the
auto-correlation  of DM particles (solid) and haloes (dashed), $\xm$
and $\xh$ respectively.  In the lower panel we plot the bias
$b^2\equiv\xh/\xm$. The dotted horizontal line marks $b=1$. Results are
shown at redshift $z=0$ (square) and 1 (triangle).  Note that, on the
scale $r\gsim 2\hmpc$, DM haloes are as biased as the matter, i.e.
$b\simeq 1$. This follows from the fact that at $z=0$ most of the
haloes have a mass $M\lsim M_\star\approx 10^{13}\mdh$ (see
Fig.~\ref{fig:haloes}), where $M_\star(z)$ is the typical mass of
haloes which collapse at redshift $z$.  The bias factor inferred from
the {\it unsmoothed}  $\xh$ and $\xm$ characterizes  the clustering of
DM  haloes.  We will propose below that it is possible using NAM to
directly measure this bias factor when the total mass is known.

\section{The action reconstruction}

The implementation of NAM used in this paper is fully described in
Peebles \etal (2001) and Phelps (2002). We refer the reader to
appendix A for details.  NAM solves for the orbits of mass tracers
given the boundary conditions that the object positions in redshift
space are fixed at the present time and that at the initial time $a^2
m_i d {\mathbf{x_i}}/dt \rightarrow 0$.  While some implementations of
NAM expand the orbits in terms of suitable trial functions (Peebles
1989, 1990, 1994; Nusser \& Branchini 2000), the present
implementation relies on matrix inversion of a discrete form of the
equations of motion (Peebles 1995, Phelps 2002), which allows the
creation of more complex orbits.  In this formulation the initial
velocities $a d {\mathbf{x_i}}/dt$ are not mathematically constrained to
vanish.  Our approach to this problem will be discussed in the
following section.

The generation of solutions begins with a set of randomized halo
orbits with 10 time steps and relaxes to a target in the gradient of
the action many orders of magnitude below that of the initial guess by
iterated per-particle matrix inversion.  Different choices of initial
randomized orbits may yield different solutions, since uniqueness is
not guaranteed, and so for each choice of cosmological parameters a
suite of 8 solutions was generated and the solution with the lowest
$\chi^2$ (see below) was used.  Orbits with negative radial distances
are allowed mathematically but in such cases the orbit is
re-randomized and the matrix inversion procedure repeated until all of
the particles have positive distances.  The resulting orbit paths
satisfy the standard equations of motion, as confirmed by a simple
leapfrog integration forward in time from the first time step of the
solution.  Solutions with intersecting orbits, which are also
mathematically allowed, are effectively suppressed by modelling each
particle as a constant density sphere with radius proportional to the
cube root of the mass, normalized so that a halo of mass $2.4\times
10^{11}\rm M_\odot$ has a radius of 180 kpc.

To more closely parallel the observational situation, where redshifts
are known to greater accuracy than the distances, the standard action
was recast with a partial transformation of coordinates, as described
in Phelps 2002, so that the redshift at $z=0$ is fixed as the boundary
condition instead of the radial distance.  The distances then emerge as
predictions, from which a standard $\chi^2$ measure is defined to
measure the goodness of fit of the solutions:
\begin{equation}
\chi^2 = \sum_i \frac{ (\mu_i^{mod} - \mu_i^{cat})^2}{\sigma_i^2},
\end{equation}
where $\mu_i^{mod}$ and $\mu_i^{cat}$ are the predicted and catalogue
distance moduli for each particle and $\sigma_i$ is the dispersion in
the observed distance moduli.  Since there are no observational errors
in the simulation, we added to the input distance moduli a gaussian
random error with a dispersion of $\sigma_i=.2$ (10 per cent distance
errors) when computing contours in $\chi^2$.
As the mean $\chi^2$ may be dominated by a few particles with
large distances errors that have settled on the wrong side of a
triple-value zone, we exclude the ten highest per-particle $\chi^2$s in
the computation of the mean.  Excluding the tail of the
distribution was found to affect the magnitude but not the
location of the minimum $\chi^2$.

Since our goal is the comparison of predicted and actual halo orbits,
it was useful to define two further quantitative measures of the
quality of the orbits, the first, ${\overline{ \Delta \theta}}$, the
``directional error'', being defined as the per-particle average angle
between the predicted and actual halo velocity vectors at $z=0$, and
the second, ${\overline{ \Delta d}}$, the ``initial displacement error'',
being the average per-particle distance between the actual and
predicted halo positions at the first time step.

\subsection{Halo mass assignment:  background mass and linear growth factor}

While each halo has a well-defined mass according to the FOF
algorithm, the total mass in haloes is only 37 per cent of the total
mass in the simulation catalogue.  If each halo were simply assigned
its bare FOF mass, then the mean density in the catalogue region as input
to NAM would be too low, and in the NAM reconstruction the region
would behave dynamically like a local void. Since the boundary
condition in the action is the redshift, and since particles at a
given distance would have higher redshifts in the reconstruction owing
to the repulsive effect of the void, NAM would predict
halo distances that are too small.

In observational catalogues, where the galaxy luminosities are given
and where $\om{m}$ is assumed to be unknown, the standard approach is
to multiply each galaxy by some constant factor, its mass to light
ratio M/L, which may be a function of galaxy type.  The minimum in
$\chi^2$ as a function of M/L will then be governed by a combination
of two factors: the degree to which the total mass in the catalogue
region matches the background density set by $\om{m}$, and the
gravitational dynamics of galaxies, groups and clusters.  In practice
the dominant contribution to $\chi^2$ is the former factor, which is
largely independent of the dynamics of the haloes themselves.

In an $N$-body simulation we define the ``halo mass multiplier''
$\mathcal{M}$, the analogue of M/L in the observational catalogues, as
the global factor by which we multiply the halo masses, which in our
simulation gives a mass density of $\om{m} = .3$ when ${\mathcal{M}} =
2.7$.  Since $\om{m}$ is a known quantity, and since it is also known
whether the region under study is close to the global mean density, it
is possible in this case to employ NAM as a probe not only of the
total mass but also of the relative distribution of mass in haloes
relative to the background, i.e. the bias.  We do this by simply
adding a smooth constant density component to the value of $\om{m}$
which is input to NAM so that, for any choice of $\mathcal{M}$, the
total mass density in the catalogue region is matched to the input
$\om{m}$.  When the total dark matter density in the catalogue region
is close to the global mass density, as it is in our simulation, this
is equivalent to multiplying $\om{m}$ as it appears in eq. A15 by the
factor ${\mathcal{M}}/2.7$.  When ${\mathcal{M}} = 2.7$ the smooth
component vanishes and all of the mass is located in the haloes as
before.  With this reassignment of the catalogue mass, a search for a
minimum in $\chi^2$ in the space of $\mathcal{M}$ becomes a test of
the linear bias relation between the underlying mass density and the
halo distribution.  To reiterate this important point: By assuming
prior knowledge of $\om{m}$, we can use NAM to probe the relative
distribution of dark matter in the haloes vs. the background.  We will
motivate this further in the following subsection.


A further modification to the halo masses is demanded by the absence
of an analytical zero velocity constraint at the initial time step.
Since in addition the haloes are modeled as non-evolving point masses
throughout their histories, NAM solutions involving close interactions
between haloes at high redshift are commonly seen.  While we have not
found a way, within the framework of the matrix inversion
implementation of the action, to write down an analytical initial-time
zero-velocity constraint, we have found that initial velocities are
effectively suppressed by the procedure, already employed in modeling
the tidal field, of rescaling the halo masses in time according to
linear perturbation theory.  We thus multiply the halo masses at each
time step by the standard linear growth factor (see Peebles 1980,
eq. 11.15), which is zero at $t=0$ and unity at $z=0$.  The value of
$\om{m}$ in eq. A15 is further rescaled by the same factor to maintain
parity at each time step with the halo mass density.  This ad hoc
procedure forces the initial velocities to vanish, eliminates hard
interactions at early times, and significantly reduces the
nonuniqueness in the solutions, but at the cost of producing halo
orbits with typical total displacements some 20 per cent shorter than
the actual paths in the simulation.  We conclude that there is room
for improvement in our method of defining halo masses, and anticipate
that some variation on the above theme, or the discovery of an
analytical zero-velocity constraint, will open the way to further
improvements in the reconstructions presented below.

\subsection{Biasing and the mass to light ratio}

Consider a discrete distribution of objects (haloes) sampling the
underlying mass density field in a region of volume $V_0$ in the
universe.  Let $n(\vx)$ be the number density of haloes and $\rho_{\rm
m}(\vx)$ be the mass density field, at comoving position $\vx$.
Denote the mean values of $n(\vx)$ and $\rho_{\rm m}(\vx)$ inside
$V_0$ by $\overline n$ and ${\overline \rho}_{\rm m}$, respectively.  Further,
choose $V_0$ to be large enough such that ${\overline \rho}_{\rm m}$ is
very close to the universal value $\om{m}\rho_{\rm c} $.  Assume that
the halo and DM density fields are related by means of a
linear bias relation between the the density contrasts $
\dh(\vx)=n(\vx)/{\overline n}-1$ and $\dm(\vx)= \rho_{\rm m}(\vx)/ {\overline
\rho}_{\rm m} -1$.  We write this biasing relation as
\begin{equation}
\dh=b\dm \; .
\label{eq:bias}
\end{equation}

We will express the bias factor $b$ in terms of the assumed $\mathcal{M}$
of haloes as it is invoked in the formalism of NAM presented
here.  This is useful in connecting this formalism to standard
applications of the linear theory of gravitational instability as well
as to other implementations of NAM (e.g. Nusser \& Branchini 2000).  A
natural way to proceed is to use the Poisson equation and the bias
relation (\ref{eq:bias}) to write the peculiar acceleration, $\vg$, of an
object in terms of $\dh$ and to compare that with the corresponding
expression derived from NAM.  The Poisson equation is
\begin{equation}
\vnabla\cdot \vg=-4\pi G{\overline \rho}_{\rm m}\dm \; ,
\label{poisson}
\end{equation}
which upon integration and substituting  $\dh=b\dm$ gives
\begin{equation}
\vg(\vx)=-G{\overline \rho}_{\rm m} b^{-1}\int\dd^3x' \dh(\vx'){\bf K}(\vx,\vx')
\label{vg1}
\end{equation}
where ${\bf K}(\vx,\vx')=(\vx-{\vx}')/|\vx-{\vx}'|^3$.
In discrete form,
\begin{equation}
\vg(\vx)=-G\frac{{\overline \rho}_{\rm m}} {{\overline n}}b^{-1}
\sum_i {\bf K}(\vx,\vx_i)+\frac{4\pi}{3}G{\overline \rho}_{\rm m} {b}^{-1}\vx
\label{vg2}
\end{equation}
NAM gives, when $a=1$,
\begin{equation}
\vg_{\rm a}=-G
\sum_i m_i {\bf K}(\vx,\vx_i)+\frac{4\pi}{3}G{\overline \rho}_{\rm mc} \vx
\label{vg3}
\end{equation}
where $m_i$ is the mass assigned to object $i$ and $ {\overline \rho}_{\rm
mc}=\sum_i m_i/V_0$ is the mean mass density estimated from the
objets.  Define $m_{0i}$ such that for $m_i=m_{0i}$ we have ${\overline
\rho}_{\rm mc}={\overline \rho}_{\rm m}$. In this case, a comparison
between (\ref{vg2}) and (\ref{vg3}) implies $m_{0i}={\overline \rho}_{\rm
m}/\overline n$ and $b=1$.  Consider now ${\overline \rho}_{\rm mc}\ne {\overline
\rho}_{\rm m}$. The difference between the mean mass densities must be
attributed to a uniformly distributed mass component which can be
associated with low (luminosity) objects that are not included in the
catalogue. Low mass objects are typically less biased than higher mass
ones (Mo \& White 1996).  Comparison between the uniform terms in
(\ref{vg2}) and (\ref{vg3}) implies
\begin{equation}
b=\frac{{\overline \rho}_{\rm m}}{{\overline \rho}_{\rm mc}} \; , \quad {\rm and}
\quad
m_i=b^{-1} m_{0i} \; .
\label{b1}
\end{equation}

In most of the applications of NAM to real data, the masses $m_i$ are
estimated from the luminosities, $L_i$, by $m_i={\mathcal{M}} L_i$
where $\mathcal{M}$ is a global mass to light ratio. Assume that
${\overline \rho}_{\rm mc}={\overline \rho}_{\rm m}$ is obtained for
${\mathcal{M}}={\mathcal{M}}_0$. 
Then, for a biased distribution of objects, the relations
(\ref{b1}) imply that the same peculiar acceleration is obtained with
the global mass to light ratio
\begin{equation}
{\mathcal{M}}=b^{-1}{\mathcal{M}}_0\; .
\label{b3} 
\end{equation}
In our simulation, since $b \simeq 1$ on the scale $r\gsim 2\hmpc$ and
since ${\mathcal{M}}_0 = 2.7$, we will expect to find the
best solutions when ${\mathcal{M}} \simeq 2.7$, i.e. when all the
mass is located in the haloes.

\section{Results}

\begin{figure}
\resizebox{0.48\textwidth}{!}{\includegraphics{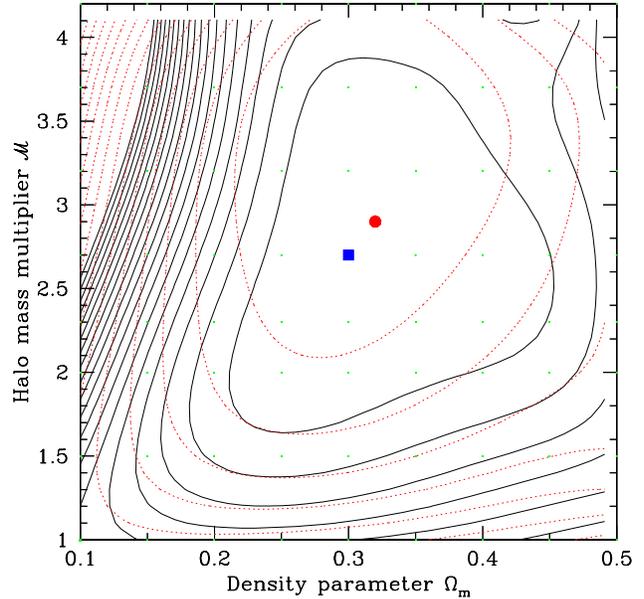}}
\caption{Mean $\chi^2$ contours showing the sensitivity of NAM to
$\om{m}$ and, through $\mathcal{M}$, to the linear bias $b$.  Solid
contours lines are from solutions including the external tidal field
and dotted contour lines are from a parallel set of solutions computed
without the tidal field.  Contour levels in both cases are .06 or
$1\sigma$.  The lowest computed mean $\chi^2$ of 1.3 (including the
tidal field) was found at $\om{m}=.3$, ${\mathcal{M}} = 2.7$
(corresponding to $b=1$).  At each grid point a set of 16 solutions
was computed, half with and half without the tidal field. For values
of ${\mathcal{M}}$ different from 2.7 a smooth component is added to
ensure that the total mass density in the catalog matches the
background density.  The filled square at center shows the global
simulation parameters, assuming a linear bias of unity, and the nearby
filled circle the parameters in the catalog region, which is slightly
overdense due to mass inflow across the catalog boundary.}
\label{fig:cont3}
\end{figure}

We searched for a minimum in $\chi^2$ in the two-dimensional parameter
space of $\om{m}$ and $\mathcal{M}$, holding $H_0$ constant at 70, by
finding NAM solutions on a grid of 56 points and interpolating a
contour map in $\chi^2$.  The results are shown in
Fig.~\ref{fig:cont3}.  At 1$\sigma$ the best solutions are found at
$.27 < \om{m} < .45$ and $1.7 < {\mathcal{M}} < 3.9$, consistent with
the simulation parameter $\om{m} = .3$ and the predicted
${\mathcal{M}} = 2.7$ (since $b \simeq 1$).  A second set of solutions
was computed without the external tidal field, and its absence had
little effect on the quality of the solutions or the location of the
minimum in $\chi^2$.  In an idealized scenario where the
exact halo masses are known, and where the region being reconstructed
is a fair sample of the global mean, NAM can thus correctly recover
both $\om{m}$ and $b$ but with an uncertainty of about 50 per cent in
both values.  In an actual observational situation with a comparable
number of particles, where significant uncertainties in the galaxy +
halo masses must additionally be taken into account, it is expected
that the uncertainties in the predicted values of $\om{m}$ and $b$
will be even larger.  It should be remarked that the tightness of
these constraints is partially a function of the size of the
catalogue, since $sigma$ in a standard $\chi^2$ measure is an inverse
square-root function of the number of particles entering into the
computation of $\chi^2$.

\begin{figure}
\resizebox{0.48\textwidth}{!}{\includegraphics{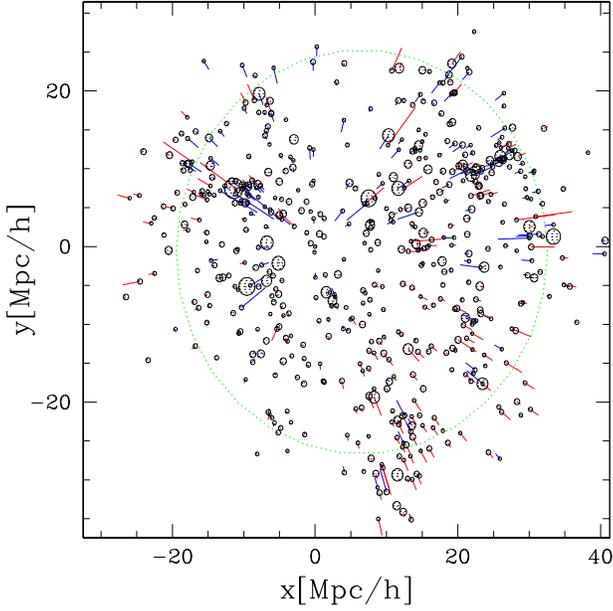}}
\caption{catalogue vs. predicted halo positions at $z=0$ in x-y projection,
for the best case solution at $\om{m}=.3$, ${\mathcal{M}} = 3.2$.
The dotted outer circle (in green) marks the nominal $26\hmpc$ boundary of the
high-density region of the simulation.
Actual halo positions at the final time step are marked by circles with radii 
proportional to $m_i^{1/3}$.  The free end of the line segments mark the
positions predicted by NAM (red when the predicted distances are larger
and blue when they are smaller); the length of the line segment thus reveals 
the radial distance error.}
\label{fig:radialerror}
\end{figure}

\begin{figure}
\resizebox{0.48\textwidth}{!}{\includegraphics{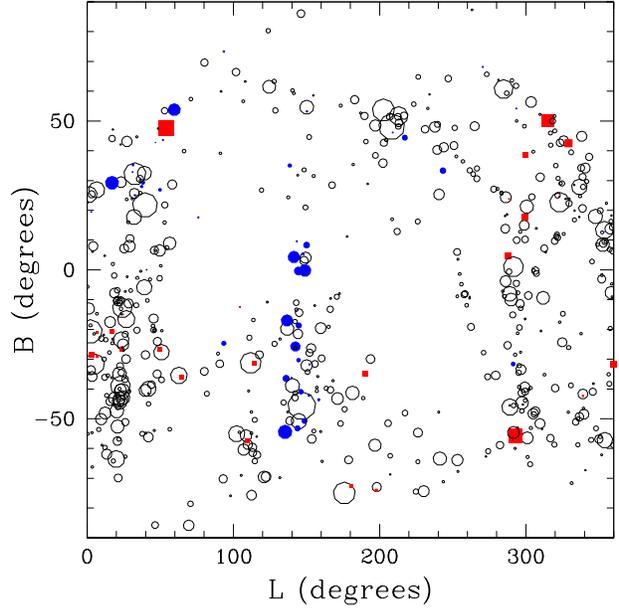}}
\caption{A view of the ``sky'' from the perspective of the reference
halo near the center of the simulation, highlighting the location of
haloes with poorly recovered distances in the best NAM solution.
Point size is proportional to the halo mass.  Filled points indicate
haloes with $\chi^2 > 1$ (radial distance error greater than 10 per
cent).  For haloes located towards the edge of the catalogue ($> 24
Mpc/h$) these are indicated by filled squares, and towards the center
of the catalogue ($< 24 Mpc/h$), by filled circles.}
\label{fig:viewfromhome}
\end{figure}

\begin{figure}
\resizebox{0.48\textwidth}{!}{\includegraphics{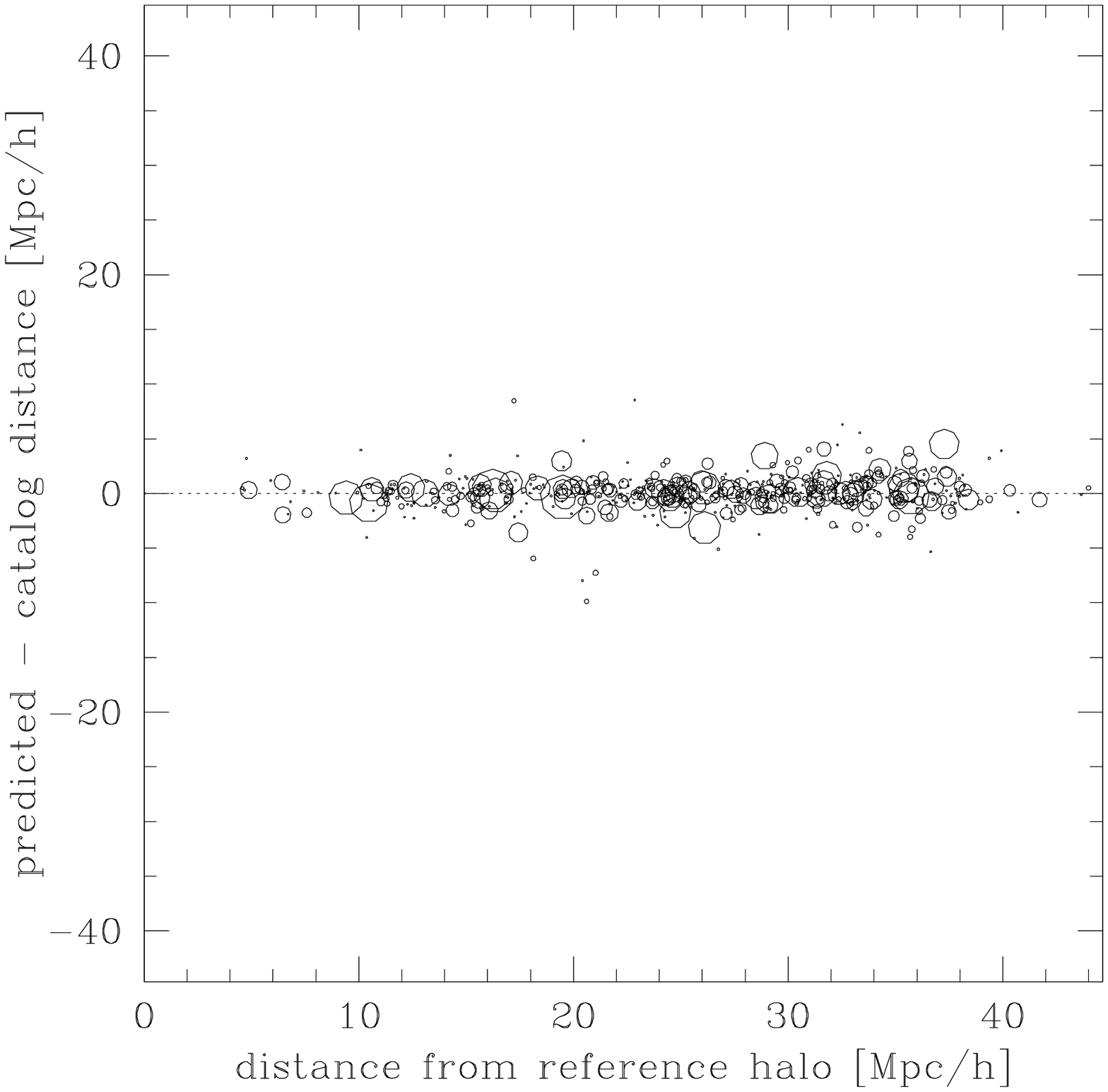}}
\resizebox{0.48\textwidth}{!}{\includegraphics{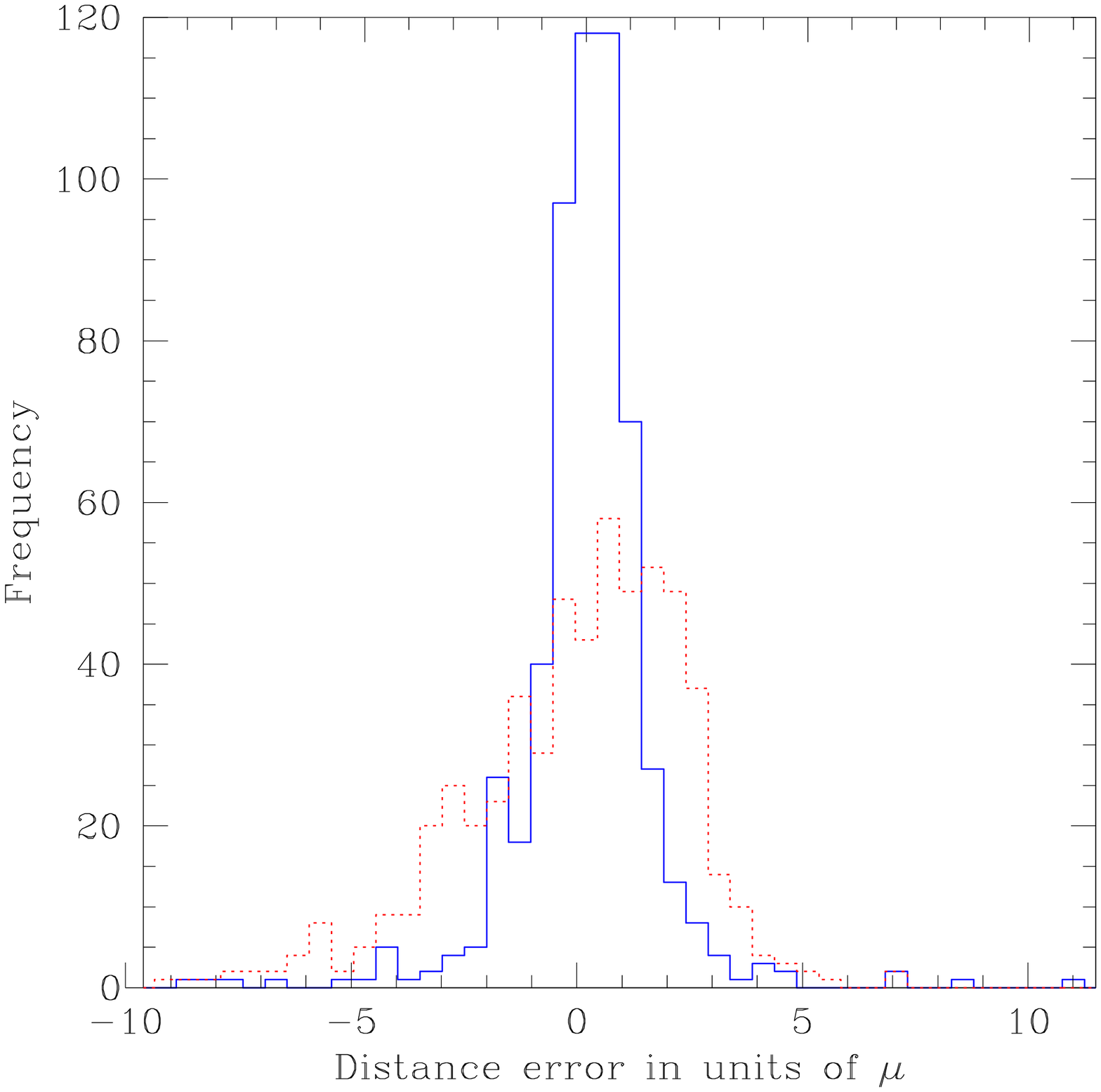}}
\caption{
{\it Top panel}~: Scatterplot of the error in radial distance predictions 
for the best NAM solution, showing good overall recovery of the distances.
Point size is proportional to halo mass.
{\it Bottom panel}~: Histogram in $\chi \equiv (\mu_i^{mod} - \mu_i^{cat})/\sigma_i$
of the same solution.  For comparison the dotted line shows the errors 
obtained when Hubble-flow distances are used in place of $\mu_i^{cat}$.}
\label{fig:scatter}
\end{figure}

\begin{figure}
\resizebox{0.48\textwidth}{!}{\includegraphics{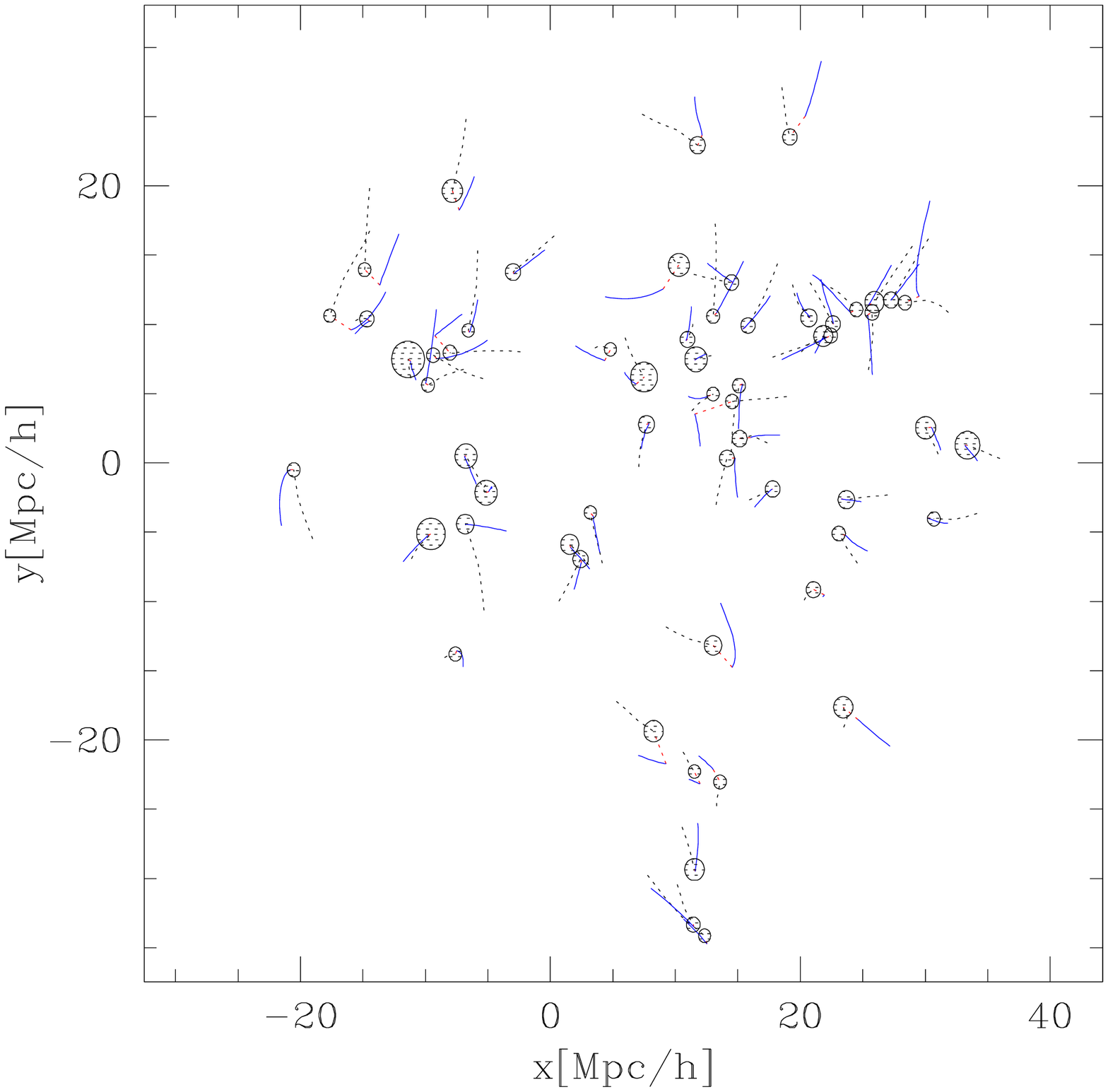}}
\resizebox{0.48\textwidth}{!}{\includegraphics{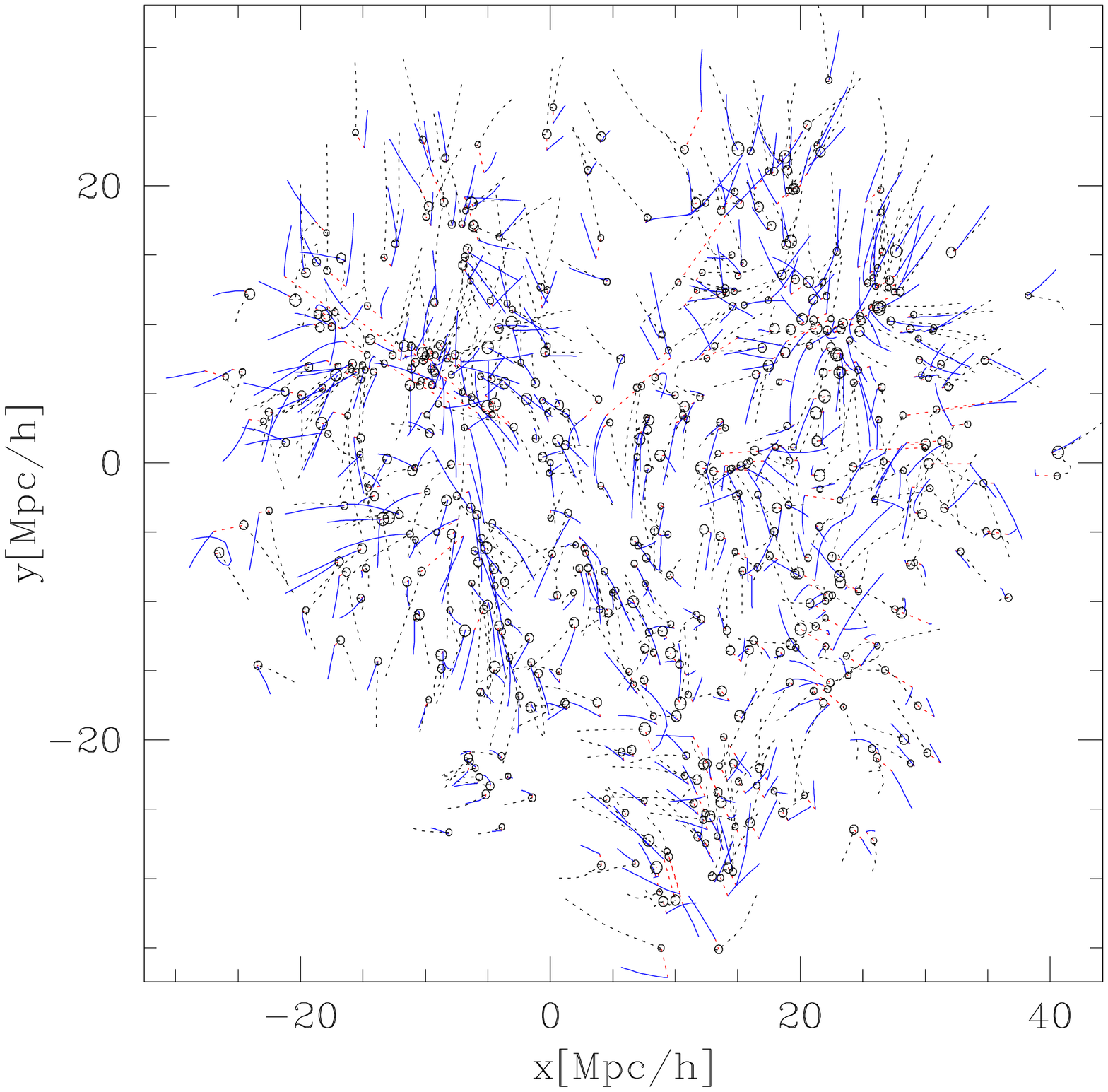}}
\caption{x-y projection of actual (solid lines) and reconstructed
orbits (dotted lines) for the heaviest haloes (top panel) and for all
other haloes (bottom panel).  Dotted lines indicate the actual
halo paths to $z=20$, while solid lines mark the paths
predicted by NAM.  Actual halo positions are indicated by circles of
radii proportional to the halo mass.  Straight radial dotted line
segments connect these positions to those predicted by NAM, as
in Fig.~\ref{fig:radialerror}.  Heavier haloes tend to have more
accurately reconstructed orbits: For the heaviest haloes $\overline {\Delta
\theta_i} = 41^o$ and for the remaining haloes $\overline {\Delta \theta_i}
= 48^o$. }
\label{fig:orbitcomp2}
\end{figure}

\begin{figure}
\resizebox{0.48\textwidth}{!}{\includegraphics{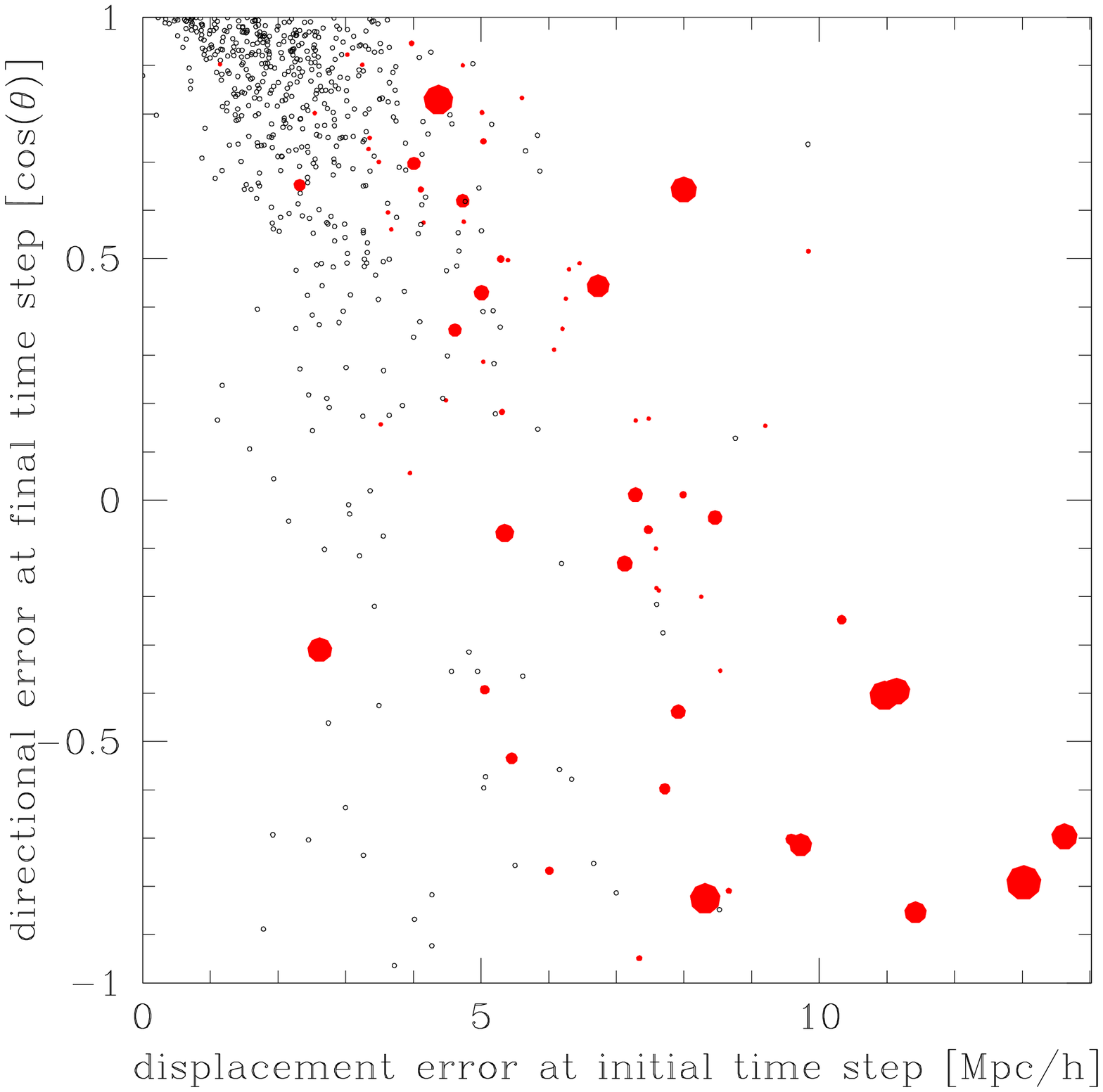}}
\caption{Scatterplot showing errors in the direction of the
reconstructed final-time velocities, as well as errors in the early
and late time predicted positions, for each of the 533 particles in
the best NAM solution.  The x-axis shows $\overline{\Delta d_i}$, the
error in the initial placement of the haloes at the first timestep.
The y-axis shows $\overline {\Delta \theta_i}$, the error in the
direction of the velocity vector at the final timestep.  Halo orbits
which most accurately predict the direction of the halo orbit at the
present epoch are thus found at the top of the graph.  Point size is
proportional $\chi^2$.}
\label{fig:all_errors}
\end{figure}

A series of closer looks at the best obtainable NAM solution
(Figs.~\ref{fig:radialerror} through \ref{fig:all_errors}), all
plotted from the same solution with input parameters $\om{m} = .3$,
${\mathcal{M}} = 2.7$, gives a good feel for the strengths and
limitations of the method.  Note that for this solution all of
the mass is located within the haloes.  Fig.~\ref{fig:radialerror}
shows in x-y projection the final positions of the haloes in the
high-density region of the simulation (open circles, with radii
proportional to the mass) and the error in the predicted positions
(line segments).  Note that the distance errors are purely radial due
to the boundary condition in the action.  The reconstruction tends to
be less accurate in the vicinity of massive haloes, which is due to
the presence of triple-value regions where NAM is easily confused.
This can be seen more clearly in Fig.~\ref{fig:viewfromhome}, which
shows a view of the ``sky'' from the reference halo, highlighting
those haloes with the largest $\chi^2$.  Some of these haloes are
found near the edge of the catalog region, where interactions with
particles in the low-resolution region, which is only approximately
modelled by NAM, may disrupt the reconstruction.  The majority
of the remaining haloes with high $\chi^2$ are often found within
a few degrees of the line of sight to a massive halo; these have
typically settled on the wrong side of a triple-value region.  Haloes
with high $\chi^2$ also tend to be found close to each other,
suggesting that if the dynamics of one influential halo are not
correctly modelled it may disrupt the accurate reconstruction of
several other nearby haloes.  Finally, a few small haloes with large
distances errors can be seen in relative isolation, indicating that
the interaction of haloes with unassociated dark matter particles, and the
complex formation and merger history of the haloes themselves, may
prevent accurate NAM reconstruction even in the absence of more
obvious disruptive factors.

The top panel of Fig.~\ref{fig:scatter} shows the error in the
predicted distances as a function of the distance from the reference
halo.  Here the excellent overall prediction of halo distances, with
typical errors of less than 3 per cent, can be seen most clearly.
Any mismatch in the halo mass density relative to $\om{m}$
would be revealed here by a tilt in the distance errors as a function
of distance from the reference halo (with a positive slope indicating
an overdensity and a negative slope an underdensity).  As the distance
errors trace a line with vanishing slope, this confirms that the total
mass for a choice of ${\mathcal{M}}=2.7$ is well matched to $\om{m}$.
The bottom panel of Fig.~\ref{fig:scatter} compares the distance
errors with those obtained assuming zero peculiar velocities and
Hubble-flow distances ($d = cz_i H_0$).  In the latter case the average
distance errors are 5 per cent.  The difference in the
sharpness of the peaks in the two histograms gives an indication of
the ability of NAM to correctly model the interparticle dynamics.

Fig.~\ref{fig:orbitcomp2} compares the NAM-reconstructed halo orbits
with the actual halo orbits, the latter being defined previously as
the CM motion of the dark matter particles comprising the halo at
$z=0$.  The reconstruction is more accurate for the heaviest haloes
(top panel) than for the rest (bottom panel): for the former the
directional error $\overline {\Delta \theta} = 41^o$, while for the latter
$\overline {\Delta \theta} = 48^o$.  By comparison, the chance
that a random orbit will have a directional error of $48^o$ or less,
that is, that a point placed at random within a sphere will fall within the
volume of the cone swept out by an opening angle of $2 * 48^o$, is about 17 per cent.
According to the directional error the quality
of the orbit reconstructions, like $\chi^2$, is not a particularly
sensitive function of ${\mathcal{M}}$: far from the $\chi^2$ minimum
at the same $\om{m}$ and ${\mathcal{M}} = 1$, $\overline {\Delta \theta_i}
\simeq 50^o$ for the entire catalogue.  Similarly, the average initial
displacement error $\overline {\Delta d_i}$, was 2.5 Mpc/h for the best
solution at ${\mathcal{M}} = 2.7$ , while far from the minimum at
${\mathcal{M}} = 1$ it was 2.9 Mpc/h.  A further indication of the
approximate character of the predicted orbits is that the magnitude of
the initial displacement error is comparable to the total distance
travelled by the typical halo orbit: The centre of mass of the average
halo in the simulation travelled 3.2 Mpc/h, while the reconstructed
haloes travelled an average of 2.6 Mpc/h, the shorter path lengths in
the reconstruction being a feature of our halo mass assignment scheme
as discussed in section 3.1.  While this error may seem large, the
chance that a random orbit with a total displacement of 2.6 Mpc/h
will end up within 2.5 Mpc/h of the actual initial position is only
16 per cent (this is the volume overlap of two spheres of equal radii 
whose centers are separated by a distance equal to their radii).
A trial NAM
reconstruction without the linear growth factor, assigning the full
halo mass at early times, gave similar values for the late-time
measures $\chi^2$ and $\overline {\Delta \theta_i}$, while the early-time
measure $\overline{\Delta d_i} \sim 4.9 Mpc/h$, or about twice the error.
The average total distance travelled by the haloes in these solutions
was 6.3 Mpc/h, nearly twice as long as the actual halo paths and
illustrative of the instability of the solutions when the haloes are
allowed to keep their full masses in the initial time steps.

Finally, Fig.~\ref{fig:all_errors} compares three measures of errors
in NAM reconstructions: $\overline{\Delta d_i}$ and $\overline {\Delta
\theta_i}$ are plotted on the x and y axes, respectively, while the
point size is proportional to $\chi^2$.  Haloes with $\chi^2 > 1$ are
shown as filled circles.  $\chi^2$ is fairly well correlated to
$\overline{\Delta d_i}$, indicating that reconstructed orbits which begin
at positions well removed from their actual starting positions in the
simulation are likely to end with inaccurately modeled distances.
Significantly, $\chi^2$ is poorly correlated to $\overline {\Delta
\theta_i}$.  This may have been expected since, in the absence of
nearby haloes constraining the dynamics, the motion of a given halo in
the plane of the sky relative to the reference halo should be fully
degenerate.  The extent to which this degeneracy is broken and the
transverse orbital motions at the present epoch are correctly
recovered is a measure of the sensitivity of NAM to the dynamics
between haloes.  The weak correlation between $\chi^2$ and $\overline
{\Delta \theta_i}$ is the clearest evidence that the predicted halo
distances alone are not a sufficient discriminator of the quality of
reconstructed orbits.

\section{Discussion and conclusions}

We have shown, using a catalogue of over 500 dark matter haloes
derived from a large $N$-body simulation at the scale of the local
supercluster, that it is possible with the numerical action method to
reconstruct the full dynamical histories of dark matter haloes given
the masses and the redshift space coordinates at the present epoch.
The reconstruction is most successful in recovering the halo distances
at the present epoch, with typical errors of less than 3 per cent.
Individual orbits paths, including the initial positions as well as
the direction of motion of the haloes at the present epoch, are
predicted with less accuracy.  By varying the relative contributions
to the total mass from the haloes and the background, we have also
found a way to use NAM to directly measure the linear bias when the
total mass density is known, although with an uncertainty of about 50
per cent.

Given the dynamical complexity of millions of interacting particles,
and the sweeping nature of NAM's central simplifying assumption that
galaxy haloes can be approximated as discrete, non-merging point
masses throughout their evolution, it is remarkable how successfully
the dynamics of a many-body system can be reconstructed on the basis
of an incomplete catalogue of facts.  The successes of NAM as it has
been implemented here are of course partially offset by their
weaknesses.  Among these is the relatively poor quality of the
reconstruction in the vicinity of massive haloes, clearly seen in
Fig.~5, which shows a breakdown in the non-linear regime where NAM,
which is itself a fully non-linear method, might have potentially
offered the most insight.  In these regions $\chi^2$ is a good
indicator of poorly reconstructed orbits, but preliminary attempts to
use this information to nudge haloes into the correct orbits while not
imposing any additional formal constraints have so far been
unsuccessful.  A second concern is the inability of NAM in many cases
to isolate, on the basis of $\chi^2$ alone, predicted halo orbits which are
moving in the wrong direction at the present epoch.  Fig.~9 shows, for
example, 34 haloes moving more than $90^o$ in the wrong direction but
with good distances at the present epoch and thus low $\chi^2$.

The above innacuracies may in part arise from
the details of our implementation, such as our ad hoc procedure of
scaling of halo masses according to linear theory, and there is
certainly room here for improvement.  The scale of the catalog is also
a factor to consider, and in particlar the density of mass tracers
within it.  This analysis should be repeated at the scale of the Local
Group, where a larger number of mass tracers acting within a smaller
volume may better constrain the dynamics and permit more accurate
orbit reconstructions.  It is also possible that, in dynamical systems
of this complexity, the angular positions, redshifts and masses are by
themselves insufficient to lift the degeneracies in the halo orbits,
and that the full three-dimensional velocities at the present epoch
will be needed to accomplish this.  This again is work to be
undertaken at Local Group scales, where next-generation observations
from SIM and GAIA hold out the promise of multiple galaxy proper
motion measurements with which to test the NAM predictions.  One
related concern is that part of the proper motion data may be needed
to recover accurate orbits, leaving fewer remaining free parameters to
assist with the more weighty problem of constraining individual galaxy
halo masses, although it is possible that only one of the two
components of the tangential velocity will be sufficient to break the
orbtial degeneracy.  Finally, inaccuracies in the NAM predictions are
doubtless due at least in part to intrinsic limitations of the method
and its assumptions, although we do not wish to suggest at this stage,
given the work that remains to be done, that that an upper
limit on NAM accuracy in orbit reconstruction has yet been reached.

We anticipate that work on NAM in the near term will lie principally
in two directions. The first is an extension of the above analysis, with
further improvements in the implementation, to a high-resolution
simulation at the scale of the Local Group, where present-day
three-dimensional velocities can provide significant additional
dynamical constraints.  The second is a direct comparison of
NAM with other reconstruction methods, both in real
space (e.g, Nusser \& Dekel 1992, Gramman 1993, Croft \& Gazta\~{n}aga
1998, Frisch \etal 2002, Mohayaee \etal 2005) and redshift space
(e.g., Narayanan \& Weinberg 1998, Monaco \& Efstathiou 1999, Mohayaee
\& Tully 2005), that help to bridge the present-day observations of
large-scale structure with the initial conditions prevailing in the
early universe.

We acknowledge the support of the Asher Space Research Institute.  We
would like to thank Felix Stoehr for providing us with the snapshots
from his simulation.

\appendix

\section{The cosmological action}

The standard cosmological action in an expanding coordinate system,
following Peebles (1989), is
\begin{eqnarray}
\mathcal{S} & = & \int_{0}^{t_0} dt \left( \sum \frac{m_ia(t)^2}{2}
\left(\frac{d\mathbf{x_i}}{dt}\right)^2 \right.  \nonumber \\ 
& & \left. + \frac{G}{a} \sum \frac{m_im_j}{\mid\mathbf{x_i}-\mathbf{x_j}\mid} \right. \nonumber \\
& & \left. - \frac{2}{3}\pi G\rho(t) a(t)^2 \sum m_i \mathbf{x_i}^2  \right),
\label{eq:action}
\end{eqnarray}
where $\mathbf{x_i}(t)$ are the co-moving coordinates, a(t) 
is the cosmological scale factor, and
$\rho(t)$ is the mean background mass density.  

The variation in the action, after integration by parts, is
\begin{eqnarray}
\delta \mathcal{S} & = & \int_0^{t_0} \dd t \, \delta \vx_i
 \left[-\frac{\dd}{\dd t}m_i a^2 \frac{\dd \vx_i}{dt} + m_i a \vg_i \right] \nonumber \\
 & & + \, \left[ m_i a^2 \delta \vx_i  \frac{\dd \vx_i}{\dd t}\right]^{t_0}_0 = 0,
\label{eq:vint}
\end{eqnarray}
where
\begin{equation}
\vg_i = \frac{G}{a^2} \sum_{j \neq i} m_j 
\frac{\vx_j - \vx_i}
{|\vx_j - \vx_i|^3} + \frac{4}{3}\pi G \rho a \vx_i.
\end{equation}
The usefulness of the action in a cosmological context hinges upon the
observation that the boundary term in (\ref{eq:vint}) vanishes either
when $\delta \vx_i = 0$ or $a^2 \dd \vx_i/\dd t = 0$.  
The former condition obtains when all three position coordinates at $t_0$
are fixed.
The latter condition is met automatically because $a = 0$ at $t = 0$.

Since cosmological redshifts can be more accurately measured than
distances, however, it is desirable to find a way to fix the observed redshifts
as a boundary condition in the action and leave the distances as free
parameters, while leaving the angular position coordinates unchanged.
The procedure can be transparently carried out in the
Hamiltonian formulation of the action, where the independent status of
generalized coordinates $q$ and momenta $p$ can be exploited:
\begin{equation}
\delta {\mathcal{S}} = \delta \int^{t_0}_0 dt \left( \sum_i p_i \dot{q}_i - H(q_i,p_i) \right) = 0.
\label{eq:ham}
\end{equation}

As described fully in Phelps (2002), the coordinate transformation
which exchanges the roles of the radial velocities and radial
distances is carried out by adding a generating function to the
Hamiltonian of the form
\begin{equation}
F = q^r p^r.
\end{equation}
where the superscript $r$ refers to the radial component. 

The modified action is
\begin{equation}
{\mathcal{S}} = \sum_i \int \left( \vec{p}_i \cdot d\vec{q}_i - 
\frac{p_i^{2}}{2m}dt
- V_i dt \right) - p_i^{r} q_i^{r}.
\end{equation}
All boundary terms are evaluated at $t = t_0$, and
\begin{equation}
V_i = -\sum_{i \neq j} \frac{Gm_im_j}{q_{ij}} 
- \sum_i \frac{\Lambda q_i^{2}}{6}.
\end{equation}
The subscripts $i$ and $j$ will refer to the particle index,
while superscripts will refer to coordinate indices.

In a homogeneous expanding coordinate system with scale
factor $a(t)$, the physical position is
\begin{equation}
a \vec{x}
\end{equation}
and the physical momentum is
\begin{equation}
m \dot{a\vec{x}} = \frac{\vec{p}}{a} + m \dot{a} \vec{x},
\end{equation}
where $\vec{x}$ is the coordinate position and where we have redefined
the variable $\vec{p} \equiv m a^2 \dot{\vec{x}}$.  Omitting for the
moment individual particle subscripts, the action is then
\begin{eqnarray}
{\mathcal{S}} & = & \int \left[ ( \frac{\vec{p}}{a}
+ m \dot{a} \vec{x} ) \cdot (a d\vec{x} + \dot{a} \vec{x} dt ) \right. \nonumber \\
 & & \left. - dt \left(\frac{\vec{p}^2}{2ma^2} +
\frac{\dot{a}}{a} \vec{p} \cdot \vec{x} + \frac{m}{2} \dot{a}^2 x^2 + V\right)
\right] \nonumber \\
 & &  \mbox{} -  \left(\frac{p^r}{a_0} + m \dot{a}_0 x^r \right) 
 a_0 x^r  \\
 & = & \int \left[ \vec{p} \cdot d\vec{x} +
ma\dot{a}\vec{x}\cdot d\vec{x}
+ \frac{m}{2} \dot{a}^2 x^2 dt \right. \nonumber \\
 & & \left. - dt \left( \frac{p^2}{2ma^2} + V \right) 
\right] \nonumber \\
 & & \mbox{} - p^r x^r - m(x^{r})^2 \dot{a}_0 a_0.
\end{eqnarray}

After integration by parts, and substituting
the standard pressureless form of the acceleration of the cosmological
expansion,
\begin{equation}
\frac{\ddot{a}}{a} = - \frac{4}{3}\pi G \rho + \frac{\Lambda}{3},
\label{eq:fried2}
\end{equation}
the action can be written as
\begin{eqnarray}
{\mathcal{S}}  =  \int \left[ \vec{p} \cdot d\vec{x} -
\left( \frac{\vec{p}^2}{2ma^2} + \tilde{V} \right)dt \right] \nonumber \\
 - p^r x^r - \frac{m (x^{r})^2 a_0\dot{a}_0}{2},
\end{eqnarray}
where $\tilde{V}$ is the gravitational potential modified by 
a smoothly distributed homogeneous
background density,
\begin{eqnarray}
\tilde{V}_i & = & - \sum_{j \neq i} \frac{G m_i m_j}{ax_{ij}} 
+ \frac{2}{3}\pi G \rho a^2 m_i x_i^2  \\
 & = & - \sum_{j \neq i} \frac{G m_i m_j}{ax_{ij}} 
+ \frac{\om{m} H_0^2}{4 a} a^2 m_i x_i^2.
\label{eq:vtilde}
\end{eqnarray}

The cosmological constant is no longer explicitly found in the action,
as expected since it cannot produce peculiar accelerations in
co-moving coordinates.  Its influence is felt in the timescale of the
integration.

The action has thus been put in the standard Hamiltonian form, with
conjugate variables $\vec{x}$ and $\vec{p}$, and with the addition of
extra terms outside the integral, the first representing the
generating function of the canonical transformation, and the second
arising from the change of boundary conditions (for further details
see Phelps 2002).

To recover the correct equations of motion, these boundary terms must
vanish in the variational derivatives, and so the constraint which was
implicit in the transformed coordinate system -- that the final-time
velocity of each particle relative to the origin is equal to its
observed redshift -- must be imposed by hand in the original
coordinates.  As the origin of the coordinate system is typically
taken to be the Milky Way, its motion along the line of sight must 
be subtracted out:
\begin{eqnarray}
v_{pec} + H_0 r - \vec{v}_0 \cdot \hat{x} & = & z, \nonumber \\
\frac{p^r}{ma_0^2} + \frac{\dot{a}_0 x^r}{a_0} - \vec{v}_0 \cdot \hat{x} 
& = & z,
\label{eq:fixz}
\end{eqnarray}
where $\vec{v}_0$ is the coordinate velocity of the Milky Way.  

With the equation of constraint (A15), the action is written
\begin{eqnarray}
{\mathcal{S}} = \frac{m}{2} a_0 \dot{a}_0 (x^{r})^2 - m a_0 z x^r  
+ m a_0^2 x^r \vec{v}_0 \cdot \hat{x} \nonumber \\ 
 \mbox{} + \int \left( \vec{p} \cdot d\vec{x} - \mathcal{H} dt \right).
\end{eqnarray}

By construction, solutions found by minimizing this modified action
will contain particle orbits whose final time-step angular positions
and radial velocities relative to the reference particle will equal
the input angular positions and redshifts.  Distances and transverse
velocities at $z=0$, as well as the orbit history of each particle,
emerge as predictions.

The rest of the problem is in principle a computational one,
and consists in devising a numerical approach to finding solutions to
the variational integral.  
The first numerical applications of the action method used a set of
fitting functions to characterize the orbits, and the action was
minimized with respect to the coefficients of these functions.  The
procedure employed here follows Peebles (1995), where the positions are
independent at each time step, allowing for the creation of sharper
orbits.

In this implementation the $\alpha^{th}$ position coordinate of the
$i^{th}$ particle at the $n^{th}$ time step is written as
$x^\alpha_{i\:n}$.  If there are $N$ time steps, then
$x^\alpha_{i\:N+1}$ is defined to represent the present-time positions
(the final time step is treated differently from the others, due to
the boundary conditions).  Time increments are defined every half
step, so that $a = 1$ at $t_{2N + 1}$ (``$t_0$'') 
and $a = 0$ at the first time step.  The momentum is then defined, as
in a standard leapfrog integration, at the half-steps, as $p^\alpha_{i
\: 2n}$.  With this notation the action integral becomes a
sum and the per-particle action is written as
\begin{eqnarray}
{\mathcal{S}}_i &  = & \frac{m_i}{2} \dot{a}_{2N+1} 
(x^r_{i \: N+1})^2 - m_i x^r_{i \: N+1} z \nonumber \\
 & & + m_i x^r_{i \: N+1} \vec{v}_{0 \: N+1} \cdot \hat{x}_{i \: N+1} \nonumber \\
 & & + \sum_{n=1}^N \left[p^\alpha_{i \: 2n} (x^\alpha_{i \: n+1} -
x^\alpha_{i \: n}) \right]  \nonumber \\
 & & - \sum_{n=1}^N \left[ \frac{(p^\alpha_{i \: 2n})^2} {2m_i a^2_{2n}}
(t_{2n+1} - t_{2n-1}) \right. \nonumber \\
 & & \left. + \tilde{V}_n(t_{2n+1}-t_{2n-1}) \right],
\end{eqnarray}
where the gravitational potential $\tilde{V}$ is defined as in
equation (\ref{eq:vtilde}).

The condition that the action be stationary with respect to the 
$p$-derivative yields an expression for $p^\alpha_{i \: 2n}$
in terms of $x$  and $t$:
\begin{equation}
\frac{\partial {\mathcal{S}}_i}{\partial p^\alpha_{i \: 2n}} = 
x^\alpha_{i \: n+1} - x^\alpha_{i \: n} -
\frac{p^\alpha_{i \: 2n}}{m_i a^2_{2n}}
(t_{2n+1}-t_{2n-1}) = 0, 
\end{equation}
\begin{equation}
p^\alpha_{i \: 2n} =  
\frac{m_i a^2_{2n}(x^\alpha_{i \: n+1} - x^\alpha_{i \: n})}
{t_{2n+1}-t_{2n-1}}.
\end{equation}

This is substituted back into the action, which is then expressed solely 
as a function of $x$:
\begin{eqnarray}
{\mathcal{S}}_i &  = & \frac{m_i}{2} \dot{a}_{2N+1} 
(x^r_{i \: N+1})^2 - m_i x^r_{i \: N+1} z  \nonumber \\
 & & + m_i x^r_{i \: N+1} \vec{v}_{0 \: N+1} \cdot \hat{x}_{i \: N+1} \nonumber \\
 & & + \sum_{n=1}^N \left[ \frac{m_i a^2_{2n}}{2}\frac{(x^\alpha_{i \: n+1} 
- x^\alpha_{i \: n})^2} {t_{2n+1}-t_{2n-1}} \right. \nonumber \\
 & & \left. - \tilde{V}_n(t_{2n+1}-t_{2n-1}) \right].
\end{eqnarray}

To find stationary points in the action through matrix inversion, by
means of the iterative approach described in Peebles (1995), the
first and second derivatives of the action with respect to the
position variables at each time step must be computed.  Calculation of
the derivatives is considerably simplified by fixing the orbits of all
particles except the one whose orbit is being solved, as the
off-diagonal terms can be ignored.  The derivatives at the final time
step are computed separately.

The gradient of the action for all time steps except the last is:
\begin{eqnarray*}
\frac{\partial {\mathcal{S}}_i}{\partial x^\alpha_{i \: n}} &  = &  
m_i a_{2n-2}^2 \frac{(x^\alpha_{i \: n}- x^\alpha_{i \: n-1})}
{t_{2n-1} - t_{2n-3}}
\end{eqnarray*}
\begin{equation}
   - m_i a_{2n}^2 \frac{(x^\alpha_{i \: n+1}- x^\alpha_{i \:n})}
{t_{2n+1} - t_{2n-1}} 
  - ({t_{2n} - t_{2n-2}}) \frac{\partial \tilde{V}_n}{\partial x^\alpha_{i \: n}} 
\end{equation}
and the second derivatives are
\begin{eqnarray*}
\frac{\partial^2 {\mathcal{S}}_i}{\partial x^{\alpha}_{i \: n}\partial 
x^{\beta}_{i \: n}} & = & m_i \delta^{\alpha \beta} \left( \frac{a_{2n}^2}{t_{2n+1} 
- t_{2n-1}} \right.
\end{eqnarray*}
\begin{equation}
\left. + \frac{a_{2n-2}^2}{t_{2n-1} - t_{2n-3}} \right) 
- (t_{2n} - t_{2n-2}) \frac{\partial^2 \tilde{V}_n}{\partial x^{\alpha}_{i \: n} 
\partial x^{\beta}_{i \:n}}
\end{equation}
for the same time step, and
\begin{equation}
\frac{\partial^2 {\mathcal{S}}_i}{\partial x^\alpha_{i \: n} 
\partial x^\alpha_{i \: n+1}} = 
- \frac{m_i a_{2n}^2}{t_{2n+1} - t_{2n-1}}
\end{equation}
for different time steps.

The gradient and second derivatives of the potential are
\begin{eqnarray*}
-\frac{\partial \tilde{V}_n}{\partial x^\alpha_{i \: n}} & = & 
 \sum_{j \neq i} \frac{Gm_i m_j}{a_{2n-1}} \frac{x^\alpha_{ji \: n}}
{x^3_{ji \: n}}
\end{eqnarray*}
\begin{equation}
+ \frac{H_0^2 \om{m}}{2 a_{2n-1}} m_i x^\alpha_{i \: n} 
\end{equation}
\begin{eqnarray}
\lefteqn {-\frac{\partial^2 \tilde{V}_n}{\partial x^\alpha_{i \: n} 
\partial x^\beta_{i \: n}}  =  \sum_{j \neq i} \frac{Gm_i m_j}{a_{2n-1}} \left[ 
 \frac{\delta^{\alpha\beta}}{{x^3_{ji \: n}}} \right. } \nonumber \\
 & & \left. - 3 \frac{x^\alpha_{ji \: n} x^\beta_{ji \: n}}{x^5_{ji \: n}}
\right] + \frac{H_0^2 \om{m}}{2 a_{2n-1}} m_i \delta^{\alpha\beta}.
\end{eqnarray}

At the final time step, when $i \neq 1$, only the components of the
gradient in the radial direction are needed:
\begin{eqnarray}
\lefteqn{\frac{\partial {\mathcal{S}}_i}{\partial x^r_{i \: N+1}}  =   
m_i (H_0 x^r_{i \: N+1} - z + \vec{v}_{0 \: N+1} \cdot \hat{x}^\alpha_{i\:N+1} )} \nonumber \\
 & & + \sum_{\alpha} m_i a_{2N}^2 \frac{\hat{x}^\alpha_{i \: N+1} \cdot (x^\alpha_{i \: N+1}
-x^\alpha_{i \: N})}{t_{2N+1} - t_{2N-1}} \nonumber \\
 & & - \sum_{\alpha} ({t_{2N+1} - t_{2N}}) \frac{\partial \tilde{V}_{N+1}}
{\partial x^\alpha_{i \: N+1}} \cdot \hat{x}^\alpha_{i \: N+1},
\end{eqnarray}
where $\hat{x}_{i\:N+1}$ is referred to the origin 
and so is purely radial.

The second derivatives are
\begin{eqnarray}
\lefteqn{ \frac{\partial^2 {\mathcal{S}}_i}
{\partial x^{r}_{i\:N+1}\partial x^{r}_{i\:N+1}}  = 
m_i H_0 + \frac{m_i a_{2N}^2}{t_{2N+1} - t_{2N-1}} } \nonumber \\
& - \sum_{\alpha} \sum_{\beta} (t_{2N+1} - t_{2N}) \frac{\partial^2 \tilde{V}_{N+1}}
{\partial x^{\alpha}_{i \: N+1}{\partial x^{\beta}_{i \: N+1}}} \nonumber \\
& \cdot \hat{x}^\alpha_{i \: N+1} \cdot \hat{x}^\beta_{i \: N+1}
\end{eqnarray}
for the same time step, and
\begin{equation}
\frac{\partial^2 {\mathcal{S}}_i}{\partial x^\alpha_{i \: N} 
\partial x^\alpha_{i \: N+1}} = 
- \frac{m_i a_{2N}^2 \hat{x}^\alpha_{i \: N+1}}{t_{2N+1} - t_{2N-1}}
\end{equation}
for the last two time steps.

\end{document}